\documentstyle[12pt,epsfig,pstricks]{article}
\catcode`\@=11
\long\def\@makefntext#1{
\protect\noindent \hbox to 3.2pt {\hskip-.9pt  
$^{{\ninerm\@thefnmark}}$\hfil}#1\hfill}		

\def\@makefnmark{\hbox to 0pt{$^{\@thefnmark}$\hss}}  
	
\def\ps@myheadings{\let\@mkboth\@gobbletwo
\def\@oddhead{\hbox{}
\rightmark\hfil\ninerm\thepage}   
\def\@oddfoot{}\def\@evenhead{\ninerm\thepage\hfil
\leftmark\hbox{}}\def\@evenfoot{}
\def\sectionmark##1{}\def\subsectionmark##1{}}

\renewcommand{\thefootnote}{\fnsymbol{footnote}}

\newcounter{sectionc}\newcounter{subsectionc}\newcounter{subsubsectionc}
\renewcommand{\section}[1] {\vspace*{0.6cm}\addtocounter{sectionc}{1} 
\setcounter{subsectionc}{0}\setcounter{subsubsectionc}{0}\noindent 
	{\normalsize\bf\thesectionc. #1}\par\vspace*{0.4cm}}
\renewcommand{\subsection}[1] {\vspace*{0.6cm}\addtocounter{subsectionc}{1} 
	\setcounter{subsubsectionc}{0}\noindent 
	{\normalsize\it\thesectionc.\thesubsectionc. #1}\par\vspace*{0.4cm}}
\renewcommand{\subsubsection}[1]
{\vspace*{0.6cm}\addtocounter{subsubsectionc}{1}
	\noindent {\normalsize\rm\thesectionc.\thesubsectionc.\thesubsubsectionc. 
	#1}\par\vspace*{0.4cm}}

\newcounter{appendixc}
\newcounter{subappendixc}[appendixc]
\newcounter{subsubappendixc}[subappendixc]

\renewcommand{\appendix}[1] {\vspace*{0.6cm}
        \refstepcounter{appendixc}
        \setcounter{figure}{0}
        \setcounter{table}{0}
        \setcounter{equation}{0}
        \renewcommand{\thefigure}{\Alph{appendixc}.\arabic{figure}}
        \renewcommand{\thetable}{\Alph{appendixc}.\arabic{table}}
        \renewcommand{\theappendixc}{\Alph{appendixc}}
        \renewcommand{\theequation}{\Alph{appendixc}.\arabic{equation}}
        \noindent{\bf Appendix \theappendixc #1}\par\vspace*{0.4cm}}

\def\abstracts#1{{
	\centering{\begin{minipage}{12.2truecm}\footnotesize\baselineskip=12pt\noindent
	\centerline{\footnotesize ABSTRACT}\vspace*{0.3cm}
	\parindent=0pt #1
	\end{minipage}}\par}} 

\renewenvironment{thebibliography}[1]
	{\begin{list}{\arabic{enumi}.}
	{\usecounter{enumi}\setlength{\parsep}{0pt}
\setlength{\leftmargin 1.25cm}{\rightmargin 0pt}
	 \setlength{\itemsep}{0pt} \settowidth
	{\labelwidth}{#1.}\sloppy}}{\end{list}}

\newcounter{itemlistc}
\newcounter{romanlistc}
\newcounter{alphlistc}
\newcounter{arabiclistc}

\newcommand{\fcaption}[1]{
        \refstepcounter{figure}
        \setbox\@tempboxa = \hbox{\footnotesize Fig.~\thefigure. #1}
        \ifdim \wd\@tempboxa > 6in
           {\begin{center}
        \parbox{6in}{\footnotesize\baselineskip=12pt Fig.~\thefigure. #1}
            \end{center}}
        \else
             {\begin{center}
             {\footnotesize Fig.~\thefigure. #1}
              \end{center}}
        \fi}

\newcommand{\tcaption}[1]{
        \refstepcounter{table}
        \setbox\@tempboxa = \hbox{\footnotesize Table~\thetable. #1}
        \ifdim \wd\@tempboxa > 6in
           {\begin{center}
        \parbox{6in}{\footnotesize\baselineskip=12pt Table~\thetable. #1}
            \end{center}}
        \else
             {\begin{center}
             {\footnotesize Table~\thetable. #1}
              \end{center}}
        \fi}

\def\@citex[#1]#2{\if@filesw\immediate\write\@auxout
	{\string\citation{#2}}\fi
\def\@citea{}\@cite{\@for\@citeb:=#2\do
	{\@citea\def\@citea{,}\@ifundefined
	{b@\@citeb}{{\bf ?}\@warning
	{Citation `\@citeb' on page \thepage \space undefined}}
	{\csname b@\@citeb\endcsname}}}{#1}}

\newif\if@cghi
\def\cite{\@cghitrue\@ifnextchar [{\@tempswatrue
	\@citex}{\@tempswafalse\@citex[]}}
\def\citelow{\@cghifalse\@ifnextchar [{\@tempswatrue
	\@citex}{\@tempswafalse\@citex[]}}
\def\@cite#1#2{{$\null^{#1}$\if@tempswa\typeout
	{IJCGA warning: optional citation argument 
	ignored: `#2'} \fi}}

 1
 1
 1

\font\ninerm=cmr9

\begin{document}
\begin{flushright}
{\bf PRA-HEP/97-13} \\
September 1997
\end{flushright}
\vspace*{1.5cm}
\centerline{\normalsize\bf PHOTOPRODUCTION OF HADRONS AT
HERA\footnote{To appear in the {\it Proceedings of the Ringberg
Workshop: New Trends in HERA Physics}, Ringberg Castle,
Germany, 25-30 May 1997.}}
\vspace*{0.6cm}
\centerline{\footnotesize JAROSLAV CVACH}
\baselineskip=13pt
\centerline{\footnotesize\it Institute of Physics, Academy of Sciences
of the Czech Republic, Na Slovance 2}
\baselineskip=12pt
\centerline{\footnotesize\it Praha, CZ - 180 40, Czech Republic}
\centerline{\footnotesize E-mail: cvach@fzu.cz}
\vspace*{0.3cm}
\centerline{\footnotesize On behalf of the H1 and ZEUS collaborations.}

\vspace*{0.9cm}
\abstracts{New experimental results from photoproduction of hadrons at HERA
are reviewed.}
 
\normalsize\baselineskip=15pt
\setcounter{footnote}{0}
\renewcommand{\thefootnote}{\alph{footnote}}
\section{Introduction}
The photoproduction of hadrons was in the first years
of HERA experimentation the process to be studied
most easily because of the large cross section.
The recent results tend to investigate the domain of
harder physics which can be compared with QCD
calculations. In this contribution the latest results
are reviewed which were presented\cite{photon97} on the spring
conferences of 1997 and are not covered by more
specialized talks of this workshop. Specifically,
I shall talk about inclusive particle production,
resonances in multi-photon final state and prompt
photons.

\section{Inclusive particle production}
The inclusive production of single hadrons in fixed
target experiments and colliding beam experiments has been one 
of the important testing grounds for the QCD parton
model. The HERA collider makes possible such
studies in $ep$ interactions over a wide
range of squared four-momentum transfer, $-Q^2$, from
photoproduction $Q^2\approx 0$ to very high
photon virtuality. 
Recent progress in the determination of parton fragmentation functions
enables us to make specific predictions for
inclusive particle production in next-to-leading
order QCD. The parton distributions of the photon
and the proton constitute an important input to 
the cross section calculations.
\newpage   
\subsection{Neutral strange particles}
New results for $K^0$ and $\Lambda$ inclusive photoproduction
were obtained in the H1 experiment\cite{h1strpho} for an average 
$\gamma p$ cms energy $W$ of $187\ \mbox{GeV}$.
$K^0_S$ mesons and $\Lambda$ baryons are identified through their decay 
channels $K^0_S \rightarrow \pi^+ \pi^-$, $\Lambda (\bar{\Lambda}) 
\rightarrow p \pi^- (\bar{p} \pi^+)$, where the pion, the proton 
and the antiproton are reconstructed in the central jet chamber (CJC).
In order to ensure an optimal acceptance of the CJC
for neutral strange particles , their pseudorapidity has been
restricted to $|\eta|<1.3$ and their transverse momentum 
to $0.5<p_t<5\ \mbox{GeV}$ for $K^0_S$ and to 
$0.6<p_t<5\ \mbox{GeV}$ for~$\Lambda$. 
\begin{figure}[h]
\begin{center}
\epsfig{file=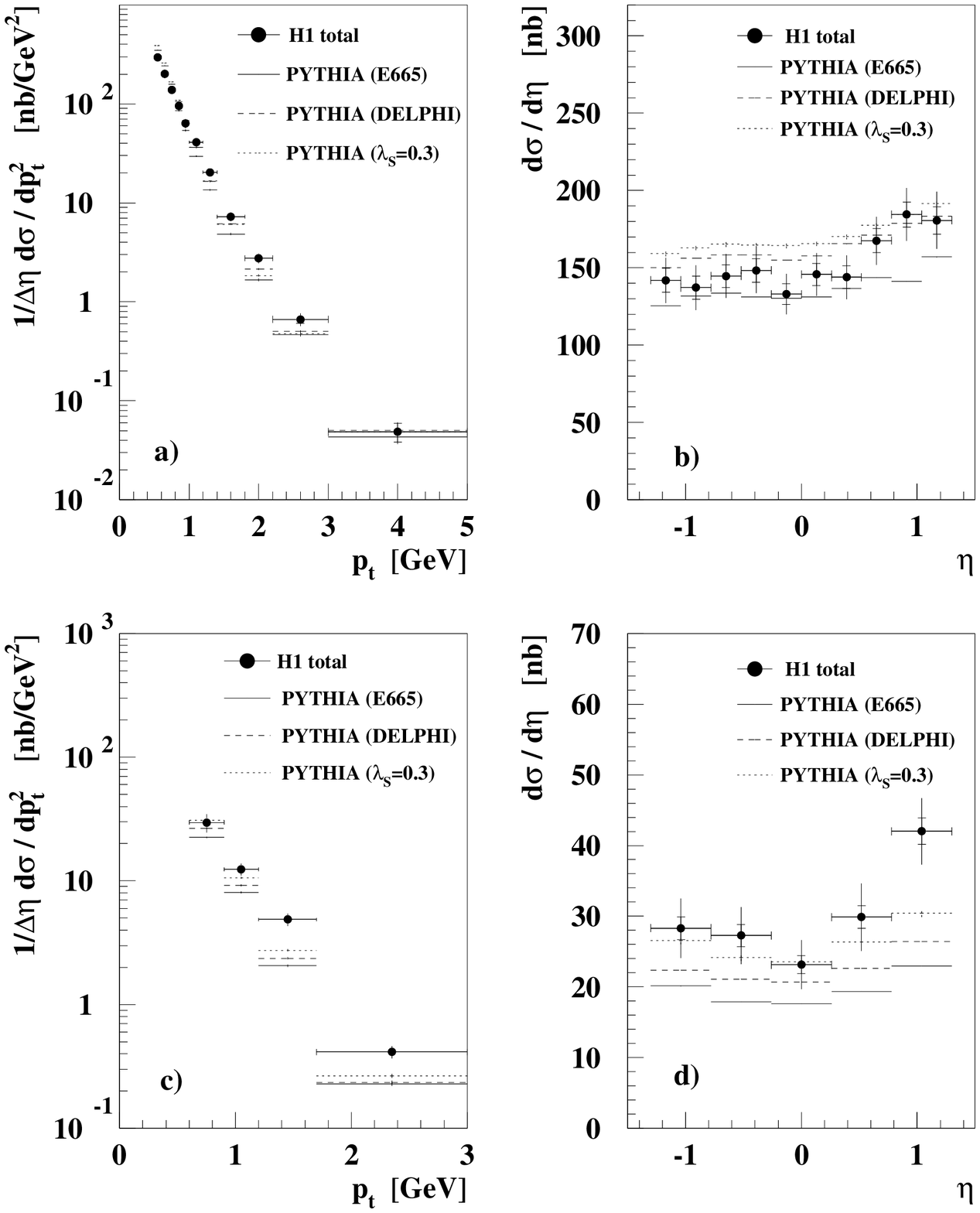,height=10cm,width=11cm}
\end{center}
\fcaption{Measured $K^0$ (a,b) and $\Lambda$ (c,d) total
$ep$ cross section in photoproduction as a function of 
$p_t$ and pseudorapidity $\eta$ compared to PYTHIA in 
the kinematic range $|\eta|<1.3$ and $0.5< p_t<5\ \mbox{GeV}$.}
\label{pythias}
\end{figure} 
All $K^0_S$ results
are multiplied by 2 and therefore correspond to
$K^0 + \bar{K^0}$ production and
are denoted by $K^0$. As the $\Lambda$ cross section 
is within errors equal to $\bar{\Lambda}$ cross section they
are combined and referenced as $\Lambda$.
Fig.~\ref{pythias} shows a comparison of the measured $K^0$ and $\Lambda$ 
cross sections in $p_t$ and $\eta$ with the predictions of 
PYTHIA in combination with JETSET using 
standard parameter settings $(\lambda_s=0.3$ as well as
parameter sets obtained from fits to DELPHI and E665
data)\footnote{The following JETSET hadronization parameters 
have been changed with respect to the default settings
(default/DELPHI/E665): $\mbox{PARJ(2)}=\lambda_s=(0.3/0.23/0.2)$,
$\mbox{PARJ(11)}=0.5/0.365/0.5)$, $\mbox{PARJ(12)}=(0.6/0.41.0.6)$.
PARJ(2) is the strangeness suppression factor,
PARJ(11) (PARJ(12)) is the probability that a meson containing $u$ or $d$ 
($s$) quark has spin 1.}.
In the case of $K^0$ mesons the measured cross sections are
in reasonable agreement with the Monte Carlo prediction
using the DELPHI or E665 settings. However the measured
$\Lambda$ cross section lies significantly above the
Monte Carlo prediction for all three settings. 
The UA5 collaboration has reported\cite{ua5} 
a similar discrepancy between 
$\Lambda$ production in $\bar{p}p$ interactions at $200\ \mbox{GeV}$
and PYTHIA predictions.

Recent NLO QCD calculations of inclusive spectra in $ep$
scattering use fragmentation functions\cite{fragf} which have been
fitted to $e^+e^-$ data. A comparison of this calculation to data
in the region of $p_t>1.8\ \mbox{GeV}$ 
is satisfactory\cite{h1strpho,kniehl}. 
If fragmentation is process independent, the rate of low 
$p_t$ particles in $\gamma p$ and deep inelastic scattering
(DIS) interactions will
be the same.
\begin{figure}[h]
\begin{center}
\epsfig{file=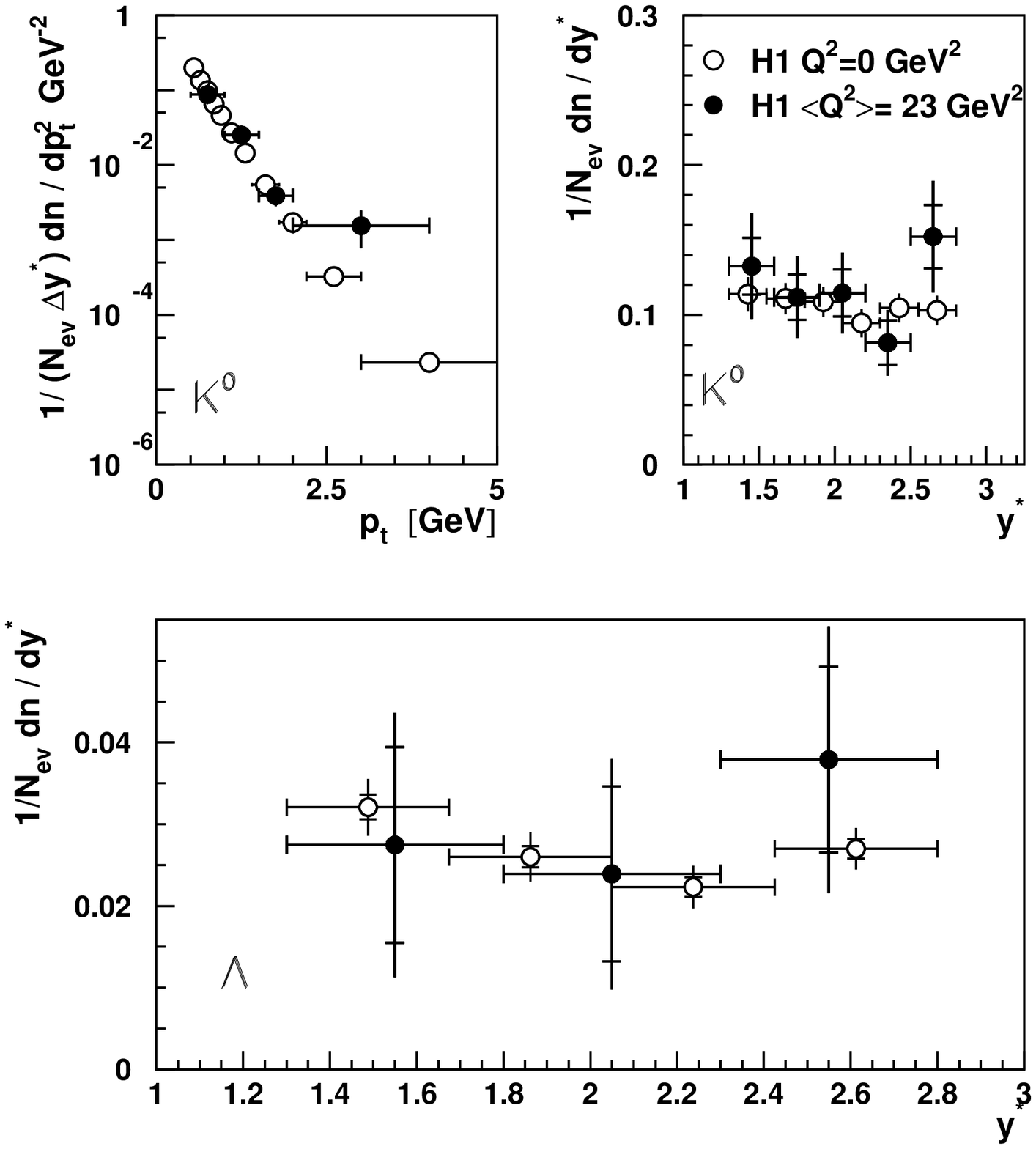,height=10cm,width=11cm}
\end{center}
\vspace{-4mm}
\fcaption{$K^0$ rate as a function of $p_t$ and centre of mass
rapidity $y^*$ (upper part) and $\Lambda$ rate as a function 
of $y^*$ (lower part) for non-diffractive events. $\gamma p$ data
are represented by open and DIS data by closed circles.
The inner vertical error bars indicate statistical error,
the outer error bars the statistical and systematic error
added in quadrature.} 
\label{klrates}
\end{figure}
A comparison of $K^0$ production rates in
$\gamma p$ and DIS\cite{h1strdis} is shown 
in Fig.~\ref{klrates} (upper part) and of $\Lambda$
production rates in Fig.~\ref{klrates} (lower part) for the
non-diffractive event samples\footnote{For non-diffractive samples 
H1 requires the presence of at least $500\ \mbox{MeV}$ of
energy in the forward region of the liquid argon calorimeter,
$2.03<\eta<3.26$, to remove large rapidity gap events.}.
The $\gamma p$ and DIS rates are compatible in $y^*$
and at low 
$p_t$ ($y^*$ is the $\gamma p$ cms rapidity defined with respect
to the direction of the exchanged boson). 
Comparison between $K^0$ rates of
production in $\bar{p}p$ collisions at 200 GeV obtained by
the UA5 collaboration\cite{ua5}  and this measurement 
show good agreement\cite{h1strpho}.
In contrast, the $\Lambda$ rates in the photon fragmentation
region in photoproduction are significantly lower than 
in $\bar{p}p$ interactions. This may be attributed to the
presence of an initial baryon in $\bar{p}p$ in this hemisphere 
from the incoming proton or antiproton. A difference between
photon and target fragmentation region with respect to baryon 
production has also been observed in deep inelastic $\mu N$
scattering\cite{e665}. 
 
ZEUS examines\cite{pa044} the features of inclusive $K^0$ mesons
produced with a high-$E_t$ jet. 
The aim of the study
is also to evaluate average longitudinal fragmentation 
function $D(z)$ for $K^0$ in jets from hard photoproduction and
compare it with the fragmentation functions\cite{fragf}
obtained in high energy quark jets
\begin{figure}[h]
\pspicture*(15.0,7.2)
\rput*[bl](1.5,0.2){
\epsfig{file=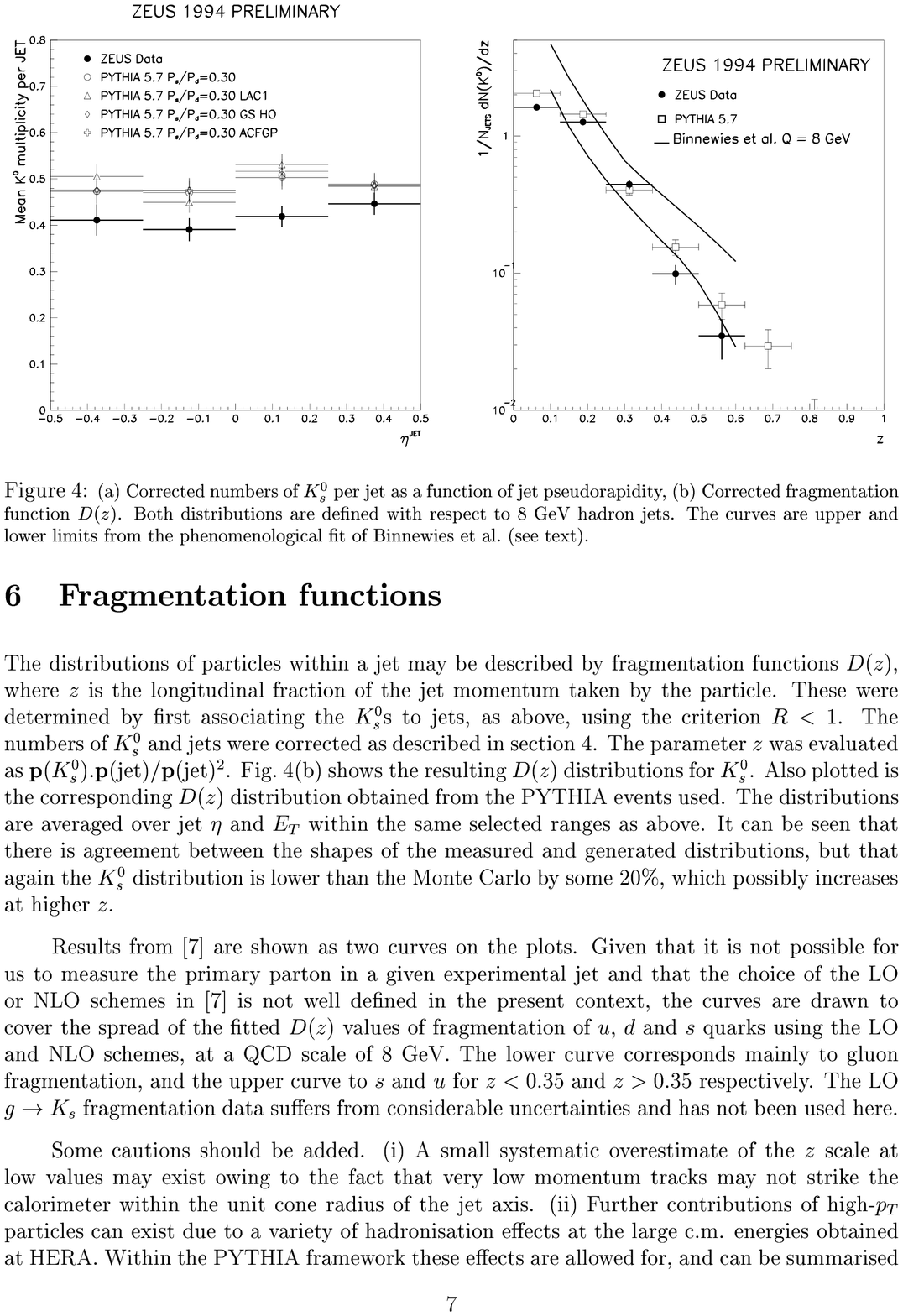,height=6cm,width=12cm,%
 bbllx=50pt,bblly=500pt,bburx=540pt,bbury=730pt,clip=}}
\rput*(3.2,1.4){(a)}
\rput*(9.4,1.4){(b)} 
\endpspicture 
\fcaption{(a) Corrected numbers of $K^0_S$ per jet as a function
of jet pseudorapidity. (b) Corrected fragmentation function $D(z)$.
Both distributions are defined with respect to $8\ \mbox{GeV}$
hadron jets.} 
\label{kfragment}
\end{figure}
in $e^+e^-$. 

Events with at least one jet of $E_t>8\ \mbox{GeV}$ and
$|\eta^{jet}|<0.5$ and $K^0_S$ with $0.5<p_t<4\ \mbox{GeV}$ 
and $|\eta|<1.5$ were selected. The data were corrected 
to take into account effects due to reconstruction efficiency,
event selection efficiency and bin-to-bin migration.
Comparison of the mean number of $K^0_S$ per jet,
integrated over $R\le 1$ ($R$ is the jet radius in cone algorithm 
in $\eta$ and $\phi$) are shown in Fig.~\ref{kfragment}a
as a function of $\eta^{jet}$ where the total number
of jets was evaluated without the $K^0_S$ requirement.
The shapes of the measured and Monte Carlo distributions
are similar but the Monte Carlo is higher than data as 
observed also in other distributions.
This discrepancy which cannot be explained by the size
of systematic errors nor variation of MC parameters
or photon parton densities is under investigation.

The distributions of particles within a jet may be
described by the fragmentation function $D(z)$, where $z$ 
is the longitudinal fraction of the jet momentum
carried by the particle. Fig.~\ref{kfragment}b shows $D(z)$
for data compared to that from PYTHIA.
Again $K^0_S$ distribution is lower than Monte Carlo by $20\%$. 
The curves are
upper and lower limits from the phenomenological fit of 
Binnewies et al.\cite{fragf} The lower curve corresponds to gluon
fragmentation and the upper curve to $s$ and $u$ fragmentations
and covers the spread of the fitted $D(z)$ values of
fragmentation of $u, d, s$ quarks using the LO and NLO
schemes at a QCD scale of 8 GeV.
The disagreement between data and phenomenological fit
of $D(z)$ can  be due to 
the fact that data are corrected to the hadron level
to avoid the strong dependence of the correction to the parton
level on the Monte Carlo used.
One should keep in mind that $e^+e^-$ data do not
provide a sufficient constraint on $g\rightarrow K^0_S$ 
fragmentation and in this place the $ep$ measurement is important.   
 
\subsection{Charged particles}
\begin{figure}[hb]
\pspicture*(15.3,8.7) 
\rput*[bl](0.,0.){\epsfig{file=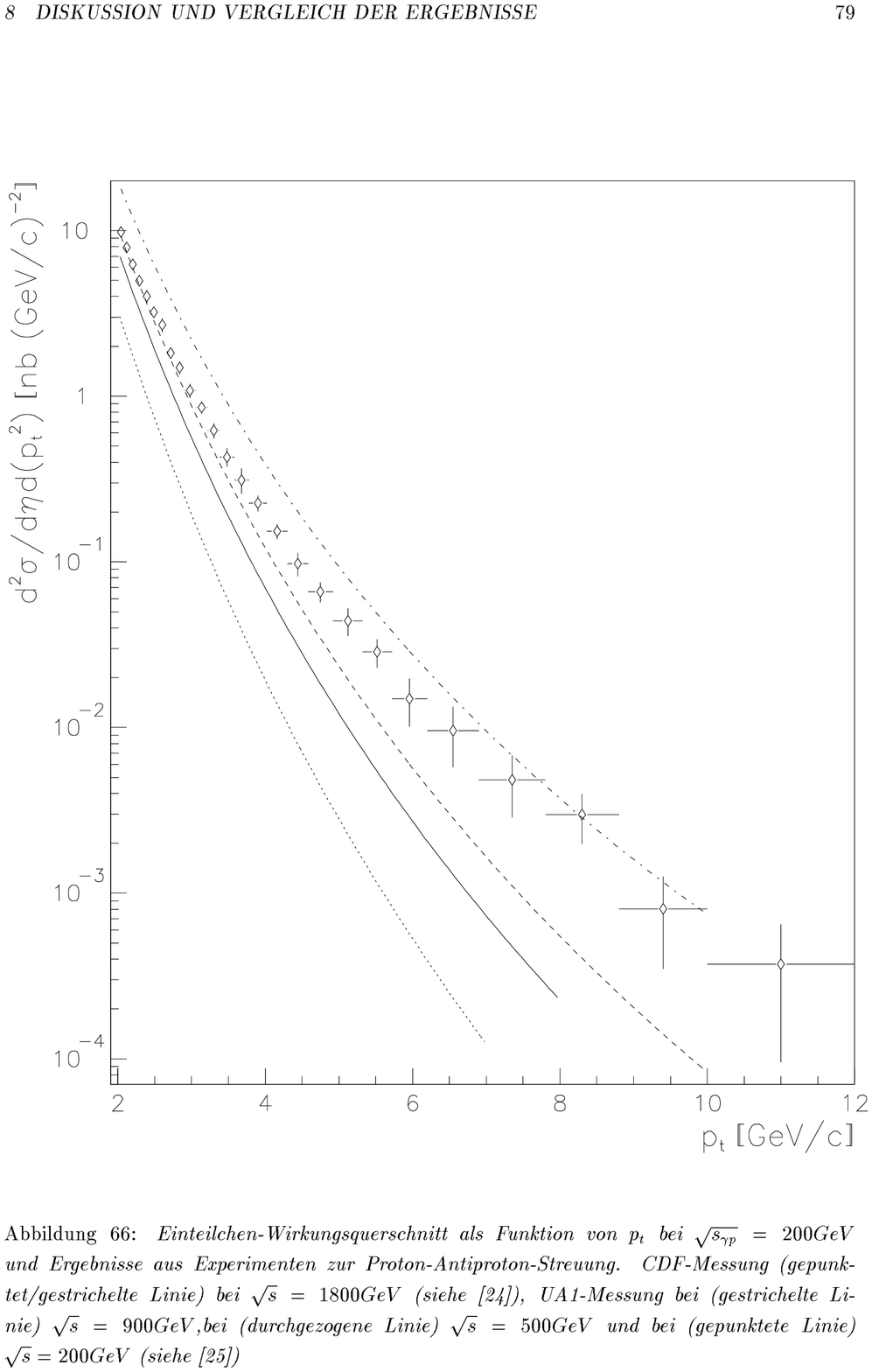,height=8.0cm,width=8.0cm,%
 bbllx=60pt,bblly=160pt,bburx=540pt,bbury=700pt,clip=} 
}
\rput*[bl](8.0,0.){   
\begin{minipage}[b]{70mm}
$p_t$ spectrum is harder in $\gamma p$ due to: 
\begin{itemize}
\item $\gamma p$ is measured in $\gamma$ fragmentation region whereas
$\bar p p$ in the central region
\item $\gamma p$ behaves like $Vp$ and parton momenta 
of quarks in the vector meson $V$ are 
harder than in baryons
\item dominance of direct part in $\gamma p$ cross section at
high $p_t$ can lead to harder scattering
\end{itemize}
\vfill
\end{minipage}
}
\rput*(5.4,4.7)
{CDF $\ 1.8\ \mbox{TeV}$}
\psbezier[linewidth=0.3pt]{->}
 (5.3,4.3)(4.9,3.3)(4.6,4.3)(4.4,3.8)
\rput*(2.1,3.2){UA1}
\rput*(2.3,2.5)
{{\small $200\ \mbox{GeV}$}} 
\rput*(2.3,2.0)
{{\small $500\ \mbox{GeV}$}} 
\rput*(2.3,1.5)
{{\small $900\ \mbox{GeV}$}}
\psline[linewidth=0.3pt]{->}(3.2,2.0)(4.0,2.5)
\psline[linewidth=0.3pt]{->}(3.2,1.5)(4.6,2.3)
\rput*(5.2,7.0){H1 preliminary}
\psdots*[dotstyle=o](3.3,7.0)
\rput*(5.1,6.2) 
{$\sqrt s_{\gamma p} = 200\ \mbox{GeV}$}
\psline[linewidth=0.5pt](0.,8.4)(15.3,8.4)
\endpspicture
\vspace*{-4mm}
\fcaption{Inclusive cross section as a function of $p_t$
for charged particles for $|\eta|<1$.
$\bar p p$ data were multiplied by $10^{-4}$ and fitted 
to the dependence $(1+\frac{p_t}{p_{0,t}})^{-n}$.
The text in the right part of the figure explains
possible reasons of the difference between 
$\gamma p$ and $\bar p p$ distributions.}
\label{ptchar}
\end{figure}

\begin{figure}[h]
\vspace*{13pt}
\rput*(5.2,4.8){(a)}
\rput*(12.0,4.8){(b)}
\def\epsfsize#1#2{0.40#1}
\epsfbox{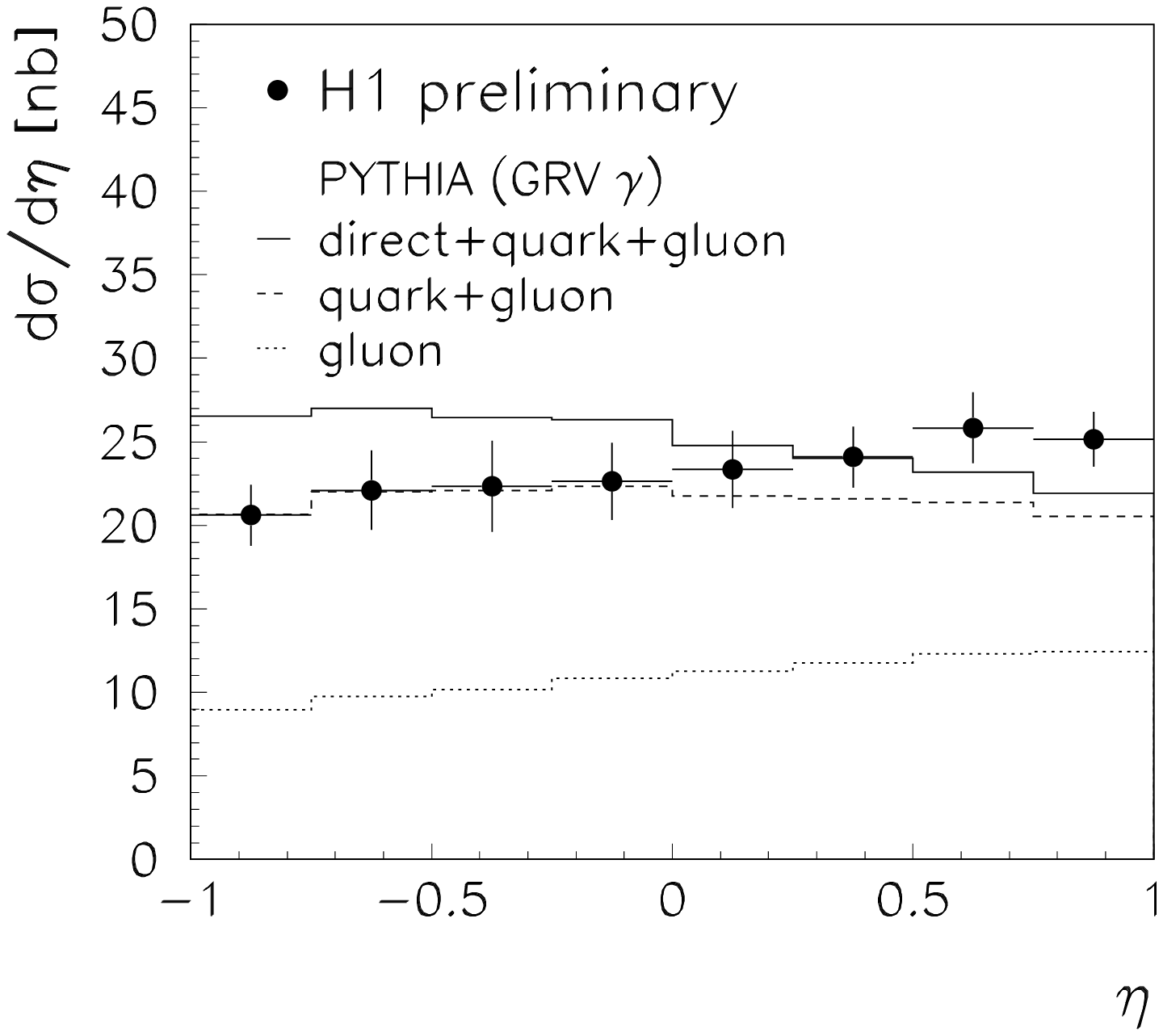}
\def\epsfsize#1#2{epsfxsize} 
\def\epsfsize#1#2{0.40#1}
\epsfbox{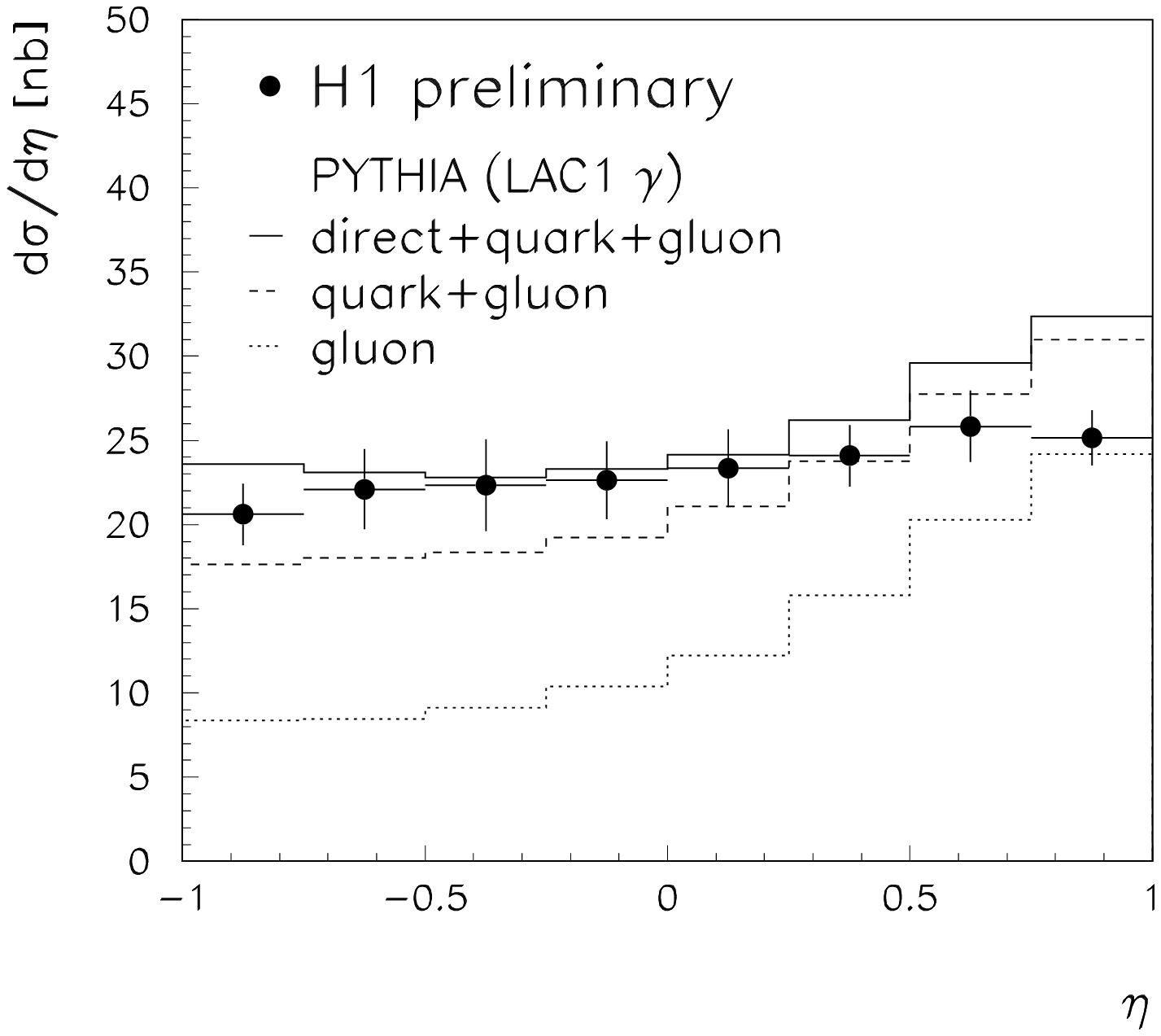}
\def\epsfsize#1#2{epsfxsize}
\fcaption{Inclusive cross section as a function of $\eta$
for charged particles with $p_t>2\ \mbox{GeV}$ compared to PYTHIA
with the GRV-LO(a) and LAC1(b) photon parton parametrisations.}
\label{etachar}
\end{figure}

Latest results of the
inclusive transverse momentum spectra of charged particles
in photoproduction events in the laboratory pseudorapidity range
$-1.2<\eta<1.4$ have been measured up to $p_t=8\ \mbox{GeV}$
using the ZEUS detector\cite{zechpho} 
and in range $|\eta|\le 1$ 
and $2<p_t<12\ \mbox{GeV}$ by H1\cite{h1chpho} 
at the average $\gamma p$
c.m. energy of $W=200\ \mbox{GeV}$. The inclusive 
transverse momentum spectra fall exponentially in the low
$p_t$ region. The non-diffractive data show a pronounced 
high $p_t$ tail departing from the exponential shape. 
Comparison to hadron--hadron collisions\cite{hadch} at a similar
c.m. energy shows that the photoproduction data are
clearly harder and in fact are similar to $\bar p p$
at $\sqrt s=1.8\ \mbox{TeV}$ (see Fig.~\ref{ptchar}).    
In Fig.~\ref{etachar}, the inclusive charged particle
cross section $d\sigma/d\eta$ is shown for particles 
with $p_t>2\ \mbox{GeV}$. A comparison is made with PYTHIA 
with two different parton distribution functions:
GRV-LO(Fig.~\ref{etachar}a) and LAC1(Fig.~\ref{etachar}b).
None of the predictions match the data
and they have mutually opposite slopes of the $\eta$ distribution.
From the figure it is clear that the rapidity
distribution is sensitive to the photon structure
function --- the fact which will be
pursued in the final analysis of the data with
the aim to extract the gluon structure function of
the photon. 
The inclusive distributions of charged particles were 
compared with NLO QCD calculations. Good agreement
is observed in $p_t$ distribution in the whole measured
range of $p_t>2\ \mbox{GeV}$ as well as in 
$d\sigma/d\eta$ distribution\cite{kniehl}. 
   
\section{Scale influence on the energy dependence of
photon-proton cross section}
\begin{figure}[h]
\pspicture*(14.0,7.9)
\rput*[bl](1.5,0.){
\epsfig{file=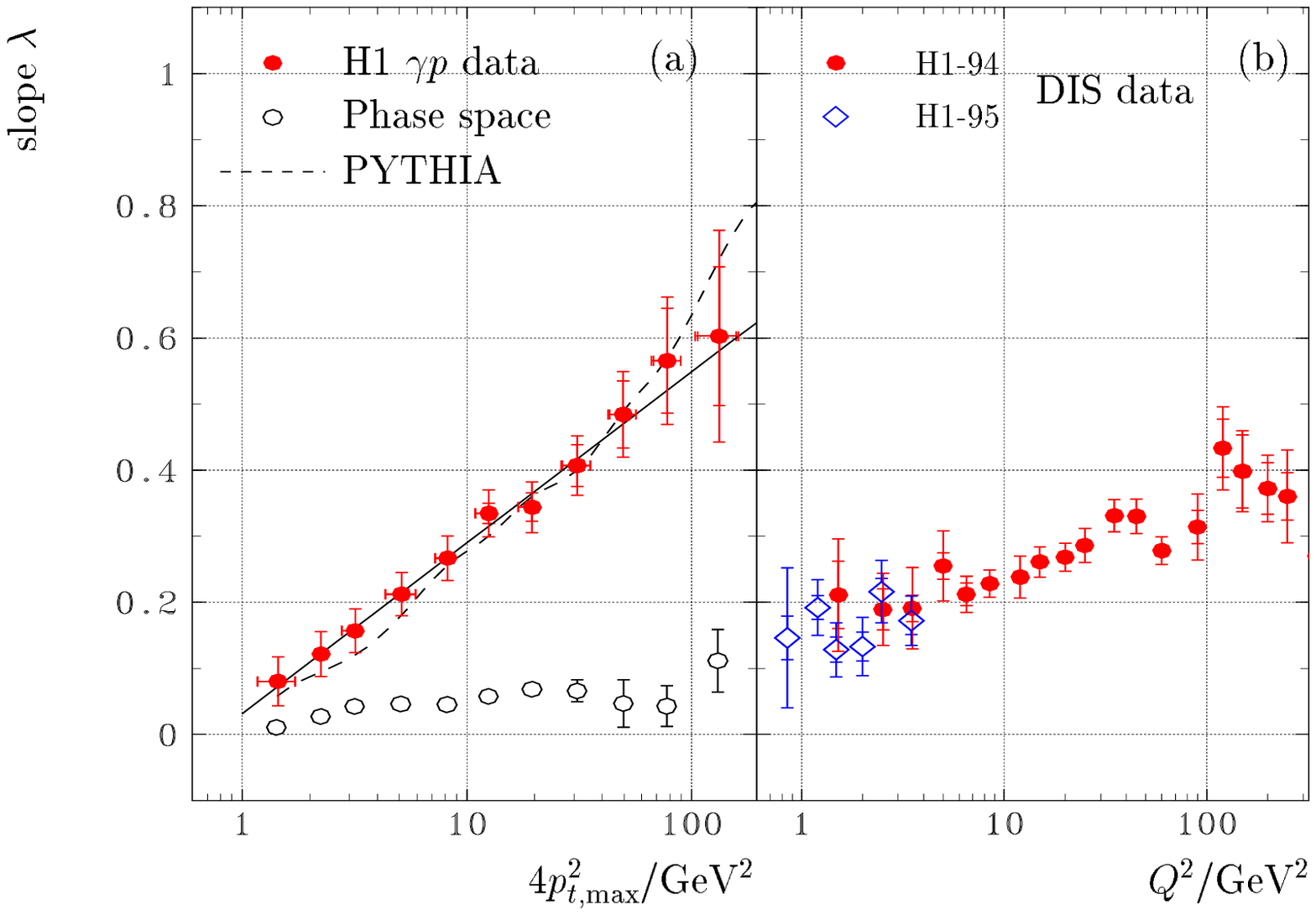,height=7.5cm,width=12cm,%
 bbllx=90pt,bblly=250pt,bburx=550pt,bbury=580pt,clip=}}
\rput*(1.6,6.6) {$\lambda$}
\endpspicture
\fcaption{Slope $\lambda$ of the $W^2$ dependence of the
photoproduction (a) and DIS (b) cross section as a function
of $4p^2_t$ and $Q^2$ respectively.}
\label{scale}
\end{figure}

The energy dependence of the real photon-proton total cross section
is consistently described at high energies by the power law
$\sigma^{tot}_{\gamma p} \propto (W^2)^\lambda$,
where $W$ is the $\gamma p$ centre of mass energy and the 
power $\lambda \sim 0.08$. In contrast, the virtual photon-proton
cross section is found to rise faster\cite{zeusf2,h1f2} 
with increasing $W^2$ 
for $Q^2$ values larger than a few GeV$^2$. In the HERA
regime fits of the form 
$\sigma^{tot}_{\gamma^* p} \propto (W^2)^\lambda$
lead to values of $\lambda$ which rise\cite{h1f2,zeusf2} 
from 0.2 to 0.4
in the $Q^2$ range from 1.5 to $10^3\ \mbox{GeV}^2$
(see Fig.~\ref{scale}).
The change of the slope parameter $\lambda$ with increasing $Q^2$
is associated with a transition from the non-perturbative `soft'
to the perturbative `hard' regime\cite{levy}.
The perturbative QCD (pQCD) can be applied in the domain 
where $Q^2$ is larger than a few GeV$^2$ and it predicts
an increase of  $\lambda$ with $Q^2$. 
In the Reggeon Field Theory (RFI) the change of $\lambda$ 
can be interpreted as a reduction of screening corrections\cite{capella} 
with increasing $Q^2$.
RFI and pQCD are two complementary
approaches which successfully describe physics processes 
in different regimes - `soft' and `hard'. 

Whereas in DIS the hardness of the
interaction (scale) can be characterized by the photon 
virtuality $Q^2$, in photoproduction with $Q^2\approx 0$
the largest $p_t$ of a hadron in the photon fragmentation
region can take over the role of the scale. 
This investigation was the subject of the study disscussed
in this chapter the results of which have been published 
recently\cite{h1scale}.

In total about two million photoproduction events were
selected in the energy range $150 < W_{\gamma p} < 250\ \mbox{GeV}$.
The $W$ dependence of the cross section was measured in different bins
of the maximal $p_{t,max}$ of the charged particle in an event.  
Good detector acceptance and high resolution of $p_t$
measurement together with large statistics of photoproduction
events, allow a continuous coverage of both soft and hard
domains in the pseudorapidity range $1.1 < \eta^* < 3.1$
($\eta^*$ is measured with respect to the $\gamma$ direction
in the $\gamma p$ centre of mass system).
The energy dependence of the cross section in each $p_{t,max}$
bin was fitted to the form $\sigma(W,p_{t,max}^2) \sim 
W^{2 \lambda(p^2_{t,max})}$. The results of
the fit are shown in Fig.~\ref{scale}, where the slope $\lambda$ is
given as function of $4p^2_{t,max}$. 
A strong increase of $\lambda$ is observed with increasing
$p_{t,max}$ which can be fitted by a straight line.
Note that when the scale $4p^2_{t,max}$ is multiplied with a number,
this does not change the slope of the straight line.
Hence the change of $\lambda$ with the scale is significant.
Comparison with the longitudinal phase space model
shows that the phase space plays only a minor role.

On the other hand the  $p_{t,max}$ dependence of $\lambda$ 
is described by PYTHIA. The model
is based on the Monte Carlo version which includes
both pQCD for the hard scattering and a Regge inspired 
soft interaction component. It was checked that the effects
of different proton and photon structure function,
of parton showers as well as multiple parton-parton
scattering are small.  Therefore, the change of the 
slope $\lambda$ is not associated with the growth of
the gluon density in the proton at small~$x$
but rather with the leading order matrix element of
the hard process in PYTHIA. 

\section{Resonances in multi-photon final state}
As mentioned already in the previous chapter the energy
dependence 
of soft photon-hadron interactions at high energies are
well described\cite{dola} 
by the Reggeon Field Theory provided
that the exchange of Pomeron trajectory with the quantum
numbers close to vacuum, i.e. with the intercept
$\alpha(t\!=\!0)\approx 1.08$ and $C$ parity $+1$ is introduced.
No direct evidence has been found at high energies for an 
exchange of trajectory with the same intercept but
with $C=-1$ called Odderon trajectory. At HERA the existence
of an Odderon trajectory could result in the production
of pseudoscalar mesons with  $C=+1$ in addition to their production
via the two photon fusion process. Sch\"afer et al.\cite{schaefer} 
suggest to measure  the cross section ratio for exclusive
production of $\eta$'s and $\eta'$'s since the
sensitivity to the Odderon contribution differs for 
$\eta$ and $\eta'$. Assuming the coupling to the meson
for photon-Odderon fusion being proportional to the product
of the electric charge and the baryon number of the quarks
composing the meson, $\eta$ - being predominantly in
the octet state with respect to flavour $SU(3)$ - gets larger
contribution compared to 
$\eta'$ being dominantly the singlet.  
\begin{figure}[h]
\begin{center}
\vspace*{13pt}
\epsfig{figure=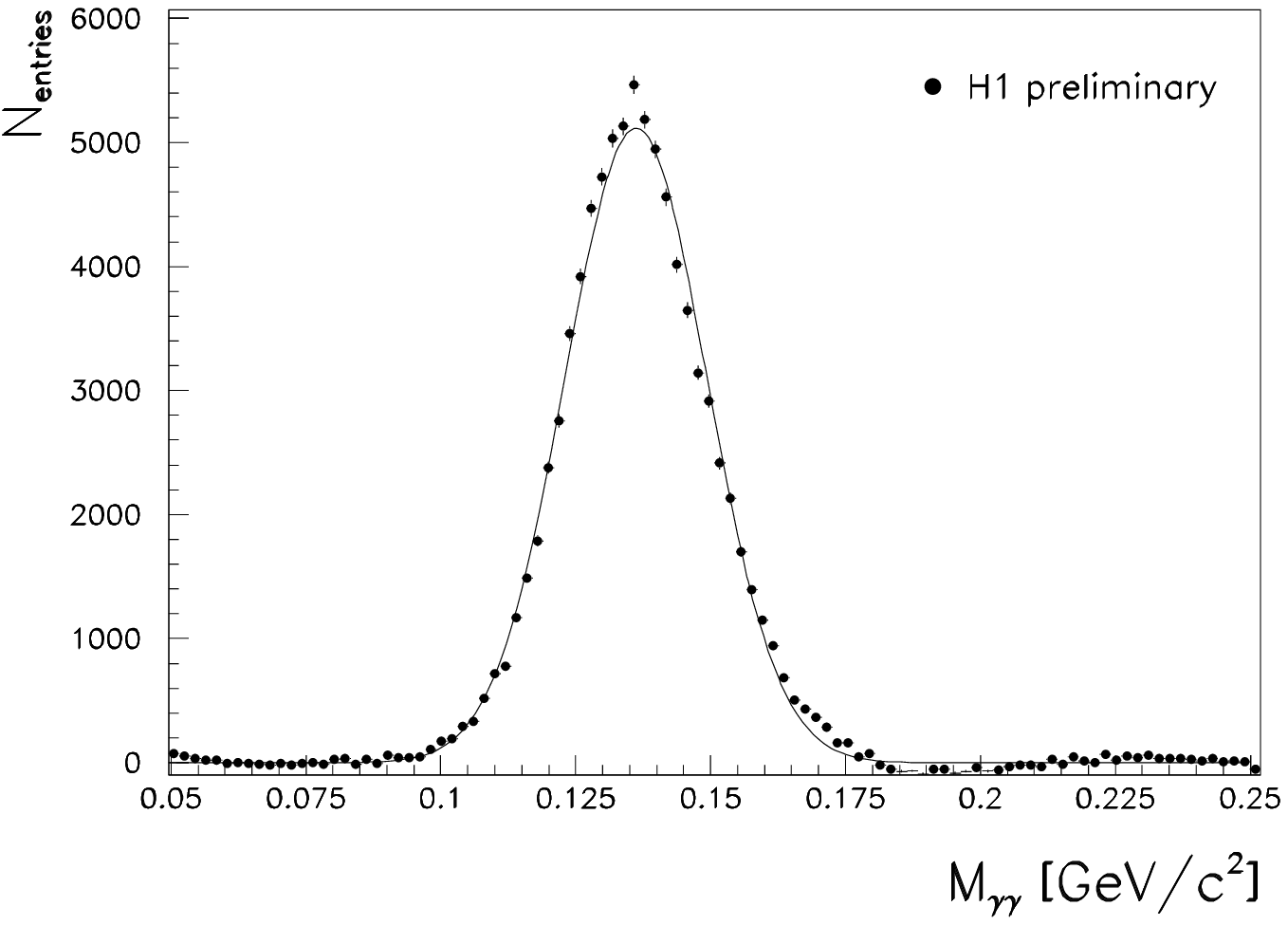,width=5.5cm} 
\rput(-4.2,3.4){\large (a)} \hspace*{1cm}
\epsfig{figure=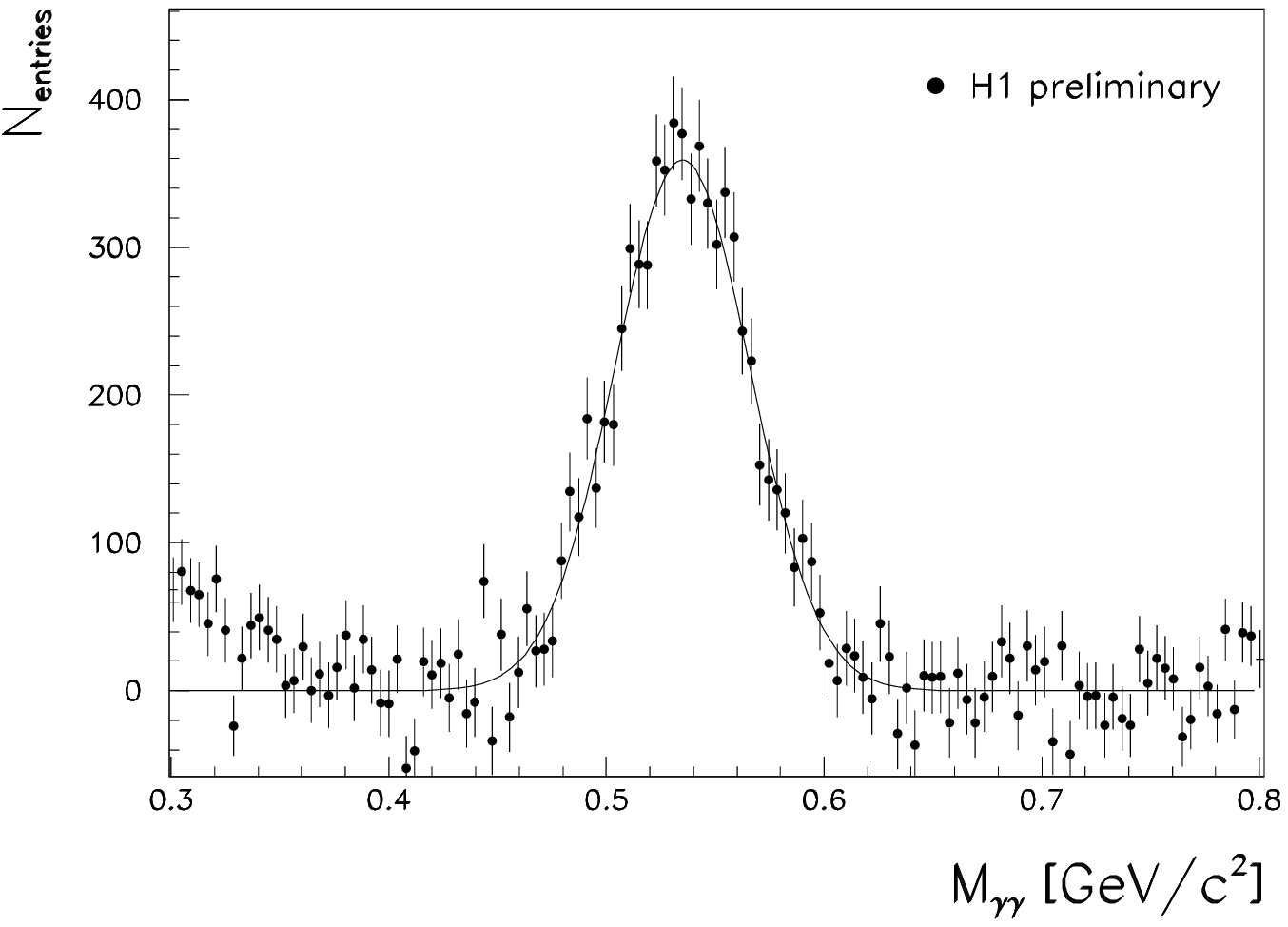,width=5.5cm} 
\rput(-4.2,3.4){\large (b)} \
\epsfig{figure=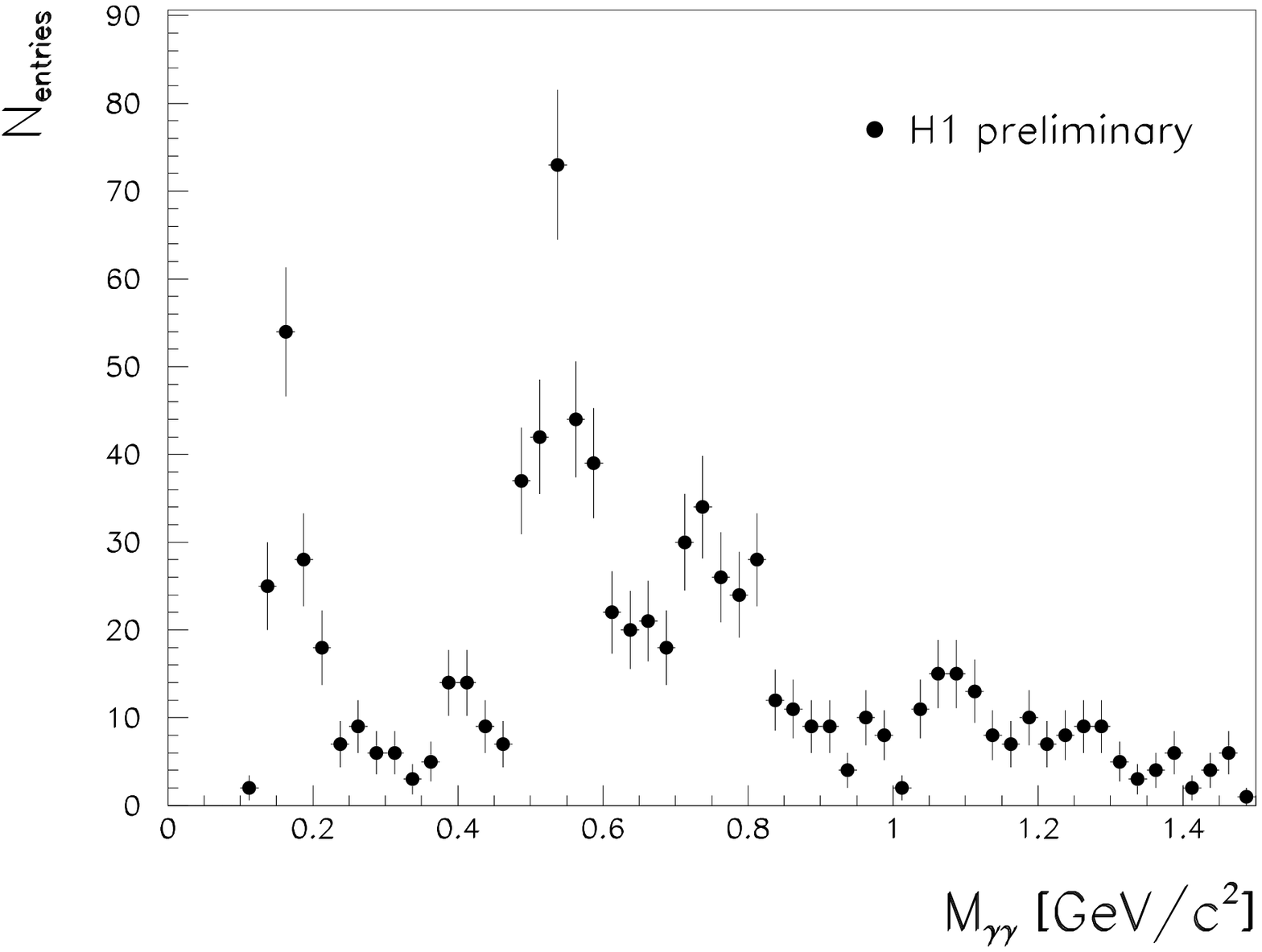,width=5.5cm} 
\rput(-4.2,3.6){\large (c)}\hspace*{1cm}
\epsfig{figure=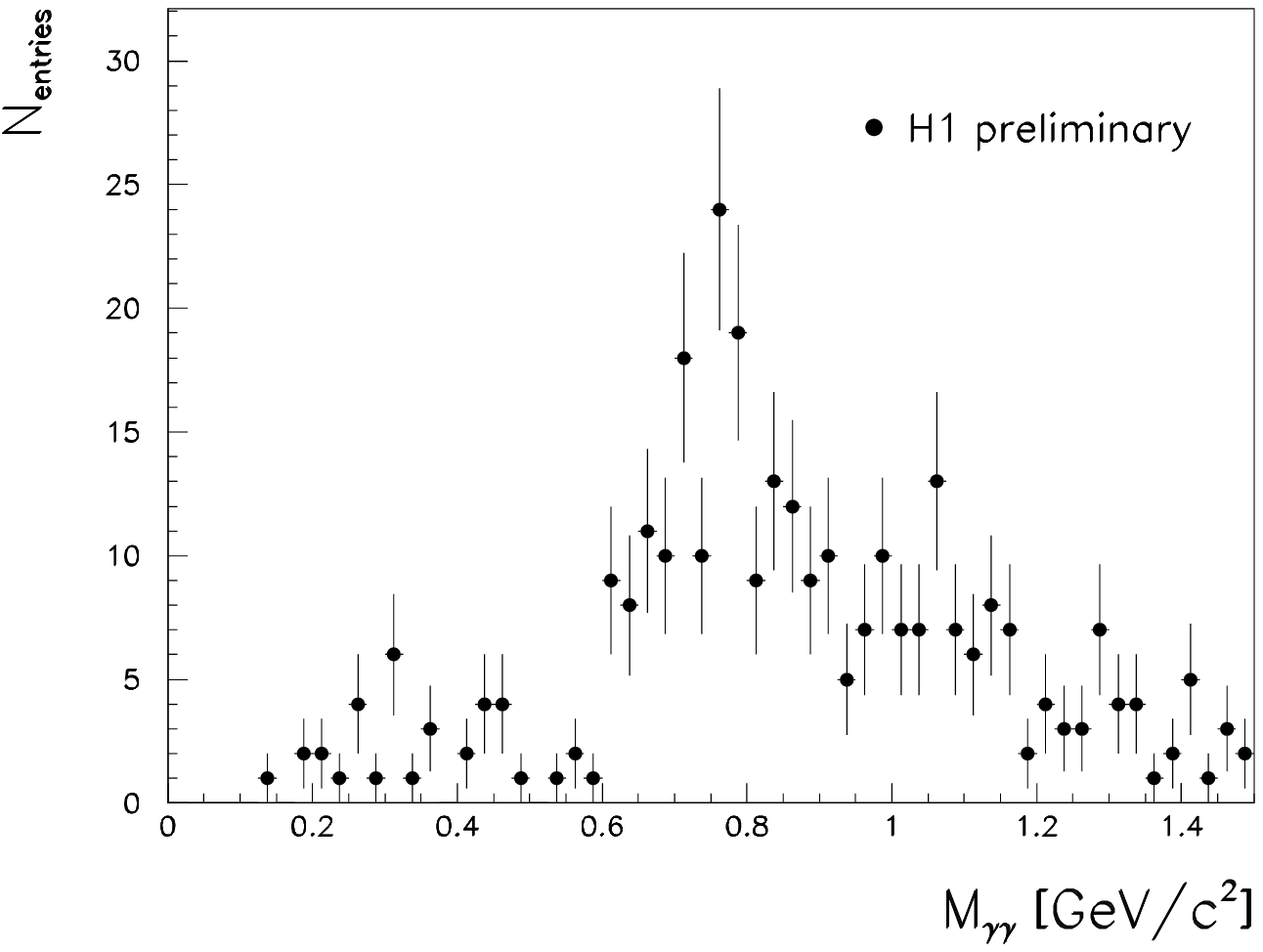,width=5.5cm}
\rput(-4.2,3.6){\large (d)}
\end{center}
\fcaption{Invariant mass spectra of $2\gamma$'s:
(a) $\pi^0$ mass region for photon
energies $E_\gamma > 1\ \mbox{GeV}$ after background subtraction 
(b) $\eta$ mass region for photon energies $E_\gamma > 1.5\ \mbox{GeV}$
after background subtraction
(c) events with a vertex and
$E_\gamma > 0.7\ \mbox{GeV}$
(d) events without a vertex and
$E_\gamma > 0.7\ \mbox{GeV}$. 
 }
\label{meson1}
\end{figure}
The data were taken in 1996 with the H1 detector and amount
to an integrated luminosity of about $5\ \mbox{pb}^{-1}$.
The lead/scintillating fibre calorimeter 
SPACAL\cite{spacal} has a sampling
term of $7\%/\sqrt{E/\mbox{GeV}}$ and a position resolution
of about $5\ \mbox{mm}$ at $1\ \mbox{GeV}$. Together with the
threshold $\approx 100\ \mbox{MeV}$ this allows for 
a precise measurement of mesons decaying into
multi-photon final states, e.g. of $\pi^0$ and $\eta$
mesons in their $2\gamma$ decay mode, as shown in Fig~\ref{meson1}. 
The central values of the Gaussian fit amount to $136\ \mbox{MeV}$
for $\pi^0$ and $535\ \mbox{MeV}$ for $\eta$.
The measured width of the invariant mass distribution is
$\sigma \approx 13\ \mbox{MeV}$ for $\pi^0$ and
$\sigma \approx 31\ \mbox{MeV}$ for $\eta$, demonstrating
the good performance of the detector.

The acceptance of the positron calorimeter of the luminosity
system restricts the photon energy in the laboratory system
to lie between 8 and 20 GeV and to 
a photon virtuality $Q^2<0.01\ \mbox{GeV}^2$. 
The selected events consist of
the scattered positron, two photons within the angular acceptance
of the SPACAL $(-3.8 < \eta^\gamma < -1.4,\  
\eta^\gamma$ is photon laboratory rapidity), 
and an elastically scattered proton which
escapes undetected in the main detector.
 
Selected events with a reconstructed vertex, i.e. with at
least one charged track in the central detector, 
result in the invariant
mass spectrum of photon pairs as shown in ~\ref{meson1}c. A clear signal
of  $\pi^0$ and $\eta$ mesons are observed, no indication for
a signal of the $\eta'$ meson at the mass 958 MeV is found. By requiring 
no charged particles in the central detector 
(events without a reconstructed vertex) is compatible with
exclusive meson production. It leads to the invariant mass spectrum
of the photon pairs as shown in ~\ref{meson1}d.
No signal of exclusively produced $\eta$ and $\eta'$ mesons
is observed nor of $\pi^0$. The resonance-like structure at
an invariant mass of about $800\ \mbox{MeV}$ is
interpreted as being due to elastic photoproduction
of $\omega$ mesons and their subsequent decays into 
the $\gamma \pi^0$ final state, where only two out of
three photons are detected.

The non-observation of $\eta$ mesons is compatible with
the expectation from a pure two photon fusion process, given the small
acceptance of the detector. This excludes a large contribution
from Odderon induced process with cross section substantially 
larger than the two photon fusion cross section.
In the case of $\eta'$ the small branching ratio
of $2.12\%$ further reduces the expected number
of events by an order of magnitude. The analysis continues to 
quantify the results into the cross section upper bounds
for exclusive production of $\pi^0$, $\eta$ and $\eta'$
mesons.

\section{Observation of isolated high $E_t$ photons at HERA.}
The ZEUS detector identified a class of events showing
the characteristics of high $E_t$ isolated photons termed
here as ``prompt'' to distinguish them from those produced
via particle decays. This is the first observation of prompt
photons at $\gamma p$ centre of mass energies an order of
magnitude higher than those previously employed.
In the kinematic range covered by  HERA detectors, 
the prompt photons are produced predominantly in two
hard subprocesses\cite{prompth} 
which in LO are represented as
the Compton scattering $\gamma q \rightarrow \gamma q$ 
in the direct channel and by $qg \rightarrow \gamma q$
in the resolved channel. NLO QCD adds additional processes
where hard photon is radiated from outgoing quarks.
Due to the electromagnetic nature of the photon
the prompt photon production is suppressed approximately
by a factor $\alpha$ with respect to hard processes
involving partons only which represent the main source
of background. Therefore, a sophisticated data
analysis must be done to isolate events with prompt
photons from background. A particular virtue of prompt photon processes 
is that the observed final-state photon emerges from
the QCD process directly, without the intermediate
hadronisation of final state partons into jets.

\begin{figure}[h]
\pspicture*(15.0,7.0)
\rput*[bl](0.5,0.0)
{\epsfig{file=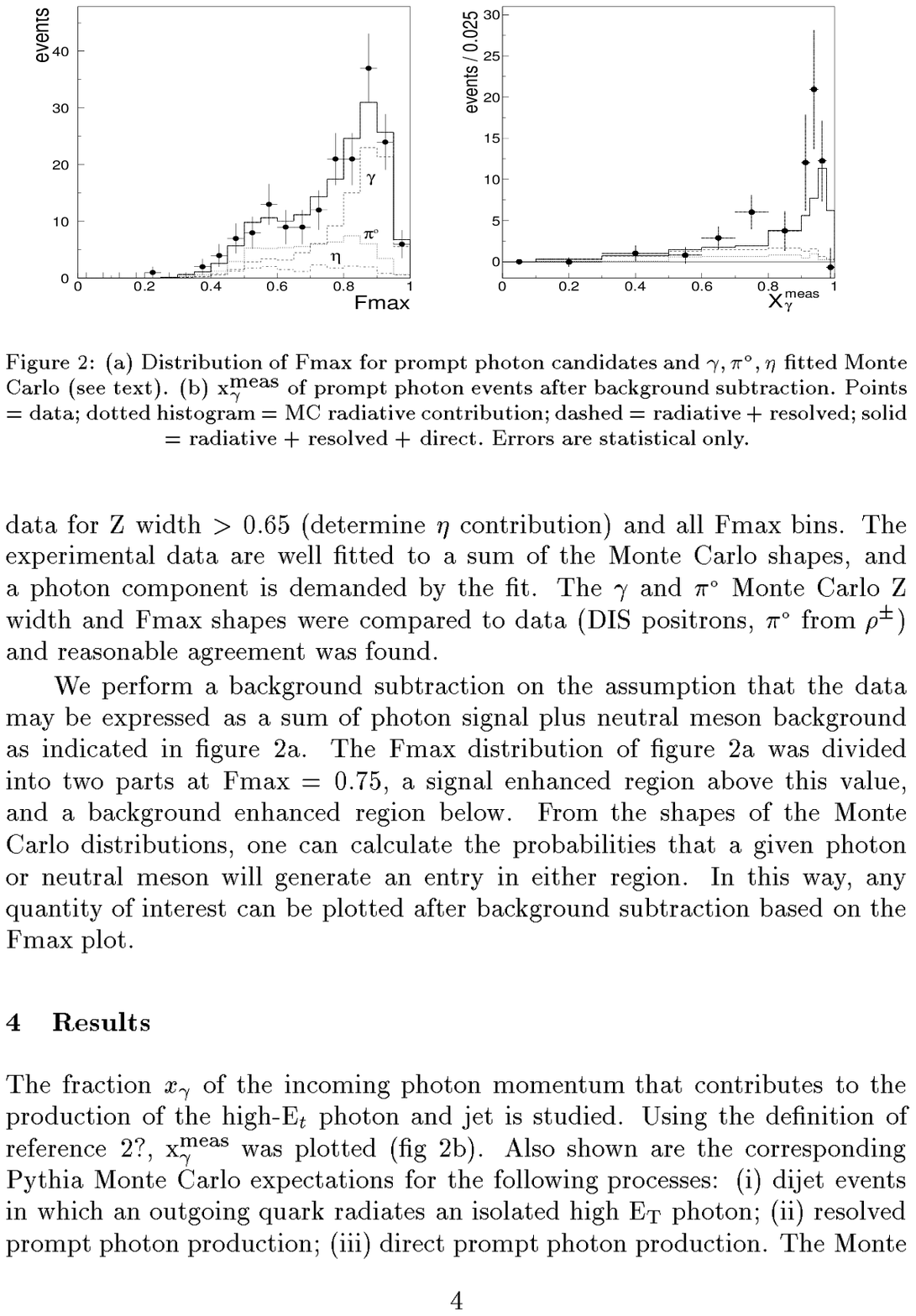,height=6.5cm,width=14cm,%
 bbllx=135pt,bblly=545pt,bburx=440pt,bbury=665pt,clip=}}
\rput*(2.2,1.8){(a)}
\rput*(3.3,5.8){$z_{width}>0.65$}
\rput*(9.6,1.8){(b)}
\endpspicture
\fcaption{(a) Distribution of $F_{max}$ for prompt photon
candidates and $\gamma,\pi^0,\eta$ fitted Monte Carlo.
$z_{width}$ given in BEMC cell units
(b) $x_\gamma ^{meas}$ of 
prompt photon after background subtraction. $(\bullet)$ data points,
histogram - Monte Carlo: $(\cdots)$ quark radiation, 
$(\_\_\_)$ q rad+resolved,
(---) q rad+resolved+direct. 
}
\label{fprompt}
\end{figure}

The data correspond to an integrated luminosity of 
$6.36\ \mbox{pb}^{-1}$ collected by the ZEUS detector 
in 1995 in $e^+p$ collisions with energy $E_e=27.5\ \mbox{GeV}$
with protons of energy $E_p=820\ \mbox{GeV}$.
Photons were detected in the barrel part of the uranium-
scintillator calorimeter in its electromagnetic section (BEMC)
which restricts the photon pseudorapidity range to 
$-0.75<\eta^\gamma<1.0$. A standard ZEUS electron finding 
algorithm was used to identify photon signal with
$5<E_t^\gamma<10\ \mbox{GeV}$. A cone jet finding
algorithm was used to identify jets with $E_t^{jet}>5\ \mbox{GeV}$, 
pseudorapidity $-1.5<\eta^{jet}<1.7$ and unit cone radius,
using entire ZEUS calorimeter system.
Events with an observed DIS electron were removed,
restricting incoming photon virtualities to $Q^2\le 1\ \mbox{GeV}^2$.
Isolation cone criteria were applied on the photon candidate
to reduce possible background from radiation of 
electrons or positrons and from energy fluctuation
of jets into hard $\pi^0$ or $\eta$.
Two shape-dependent quantities were studied in order
to distinguish photon, $\pi^0$ and $\eta$ signals.
These were the energy weighted mean width of the BEMC cluster
in z direction $z_{width}$ and the energy fraction
of the most energetic cell in the cluster 
$F_{max}$ (Fig.~\ref{fprompt}a) of the photon candidate.
The relative amounts of $\gamma$, $\pi^0$ and $\eta$ from
single particle Monte Carlo distributions were obtained from fits of 
$z_{width}$ and $F_{max}$ to data.
To get distribution of physical quantities,
a background subtraction is applied  on the assumption that 
the data may be expressed as a sum of the photon signal plus 
neutral meson background. 
The result for the fraction $x_\gamma$ of the incoming photon momentum that 
contributes to the production of the high $E_t$ photon and jet
$x_\gamma^{meas}=\sum_{jet+\gamma} (E-p_z)/ \sum_{event} (E-p_z) $ is 
shown in Fig.~\ref{fprompt}b. The Monte Carlo distributions
are normalised to the same integrated luminosity as data.
A strong peak near $x_\gamma^{meas}=1$ corresponds
to the Compton process. It is not possible to make conclusions
concerning the presence of a resolved photon component.
The background distribution is consistent
with zero for $x_\gamma^{meas}>0.9$.
The quoted cross section for the process
$e p \ \rightarrow \ e+\gamma_{prompt}+ jet + X$
with $x_\gamma^{meas}\ge 0.8$ is 
$15.3\pm3.8\mbox{(stat)}\pm1.8\mbox{(syst)}\ \mbox{pb}$
in the kinematic range defined above.
The systematic error is dominated by $8\%$ contribution due
to the uncertainty in the calorimeter calibration.
The cross section is in an agreement with the NLO
QCD calculation\cite{gordon}.

\section{Conclusion} 
Inclusive particle production provides a complementary view on hard scattering 
to the commonly more used jet production. 
Selecting events with a hadron with
$p_t$ at least $1-2\ \mbox{GeV}$, the inclusive spectra
are calculable in QCD and are sensitive to structure
functions of beam projectiles --- photon (real or virtual) and
proton.  It will be the task of the coming years of HERA 
hadron physics to use inclusive spectra of hard hadrons
to obtain the gluon structure function of the real photon.
Events with prompt photons have a sensitivity to the
quark content of the photon. With the increase of HERA
luminosity they will supplement
the inclusive spectra of charged particles in
the study of partonic structure of the photon.

\section{Acknowledgement}
It is a pleasure for me to thank my colleagues from ZEUS and H1
collaboration who helped to prepare this talk and provided me
with results and figures. I would also like to thank the
organizers of the workshop for a very friendly atmosphere
and a place for many stimulating discussions during
the workshop.

This work is supported by the Grant Agency of Academy of 
Sciences of the Czech Republic under the grant no. A1010619
and by the Grant Agency of the Czech Republic under the 
grant no. 202/96/0214.


\end{document}